\documentclass[10pt,journal]{IEEEtran}

\ifCLASSINFOpdf
\else
\fi
\usepackage{hhline, array, ragged2e, multirow, tabularray, booktabs}
\usepackage{graphicx, amssymb, comment}
\usepackage{amsmath, float, tabularx, colortbl, arydshln}
\usepackage[normalem]{ulem}
\usepackage[noabbrev]{cleveref}
\usepackage{algorithm}
\usepackage{algpseudocode}

\usepackage{verbatim}

\renewcommand{\algorithmicensure}{\textbf{Function:}}

\begin{document}
\title{Robustness Analysis against Adversarial Patch Attacks in Fully Unmanned Stores}

\author{\IEEEauthorblockN{Hyunsik Na, Wonho Lee, Seungdeok Roh, Sohee Park and Daeseon Choi}

\thanks{Manuscript received January 31. (Corresponding author: Daeseon Choi.)}
\thanks{This work was partially supported by the Institute of Information and Communications Technology Planning and Evaluation (IITP) grant through the Korean Government [Ministry of Science and ICT (MSIT)] (Robust AI and Distributed Attack Detection for Edge AI Security) under Grant 2021-0-00511 and by the IITP grant funded by the Korea government (MSIT) (No. RS-2024-00398353, Development of Countermeasure Technologies for Generative AI Security Threats).}
\thanks{H. Na, W. Lee, S. Roh and S. Park are with Department of Software Convergence, Soongsil University, Seoul 07027, South Korea, Email: rnrud7932@soongsil.ac.kr.}
\thanks{D. Choi is with Department of Software, Soongsil University, Seoul 07027, South Korea, Email: sunchoi@ssu.ac.kr.}}

\markboth{Journal of \LaTeX\ Class Files,~Vol.~14, No.~8, August~2015}%
{Shell \MakeLowercase{\textit{et al.}}: Bare Demo of IEEEtran.cls for IEEE Transactions on Magnetics Journals}

\IEEEtitleabstractindextext{%
\begin{abstract}
The advent of convenient and efficient fully unmanned stores equipped with artificial intelligence-based automated checkout systems marks a new era in retail. However, these systems have inherent artificial intelligence security vulnerabilities, which are exploited via adversarial patch attacks, particularly in physical environments. This study demonstrated that adversarial patches can severely disrupt object detection models used in unmanned stores, leading to issues such as theft, inventory discrepancies, and interference. We investigated three types of adversarial patch attacks---Hiding, Creating, and Altering attacks---and highlighted their effectiveness. We also introduce the novel color histogram similarity loss function by leveraging attacker knowledge of the color information of a target class object. Besides the traditional confusion-matrix-based attack success rate, we introduce a new bounding-boxes-based metric to analyze the practical impact of these attacks. Starting with attacks on object detection models trained on snack and fruit datasets in a digital environment, we evaluated the effectiveness of adversarial patches in a physical testbed that mimicked a real unmanned store with RGB cameras and realistic conditions. Furthermore, we assessed the robustness of these attacks in black-box scenarios, demonstrating that shadow attacks can enhance success rates of attacks even without direct access to model parameters. Our study underscores the necessity for robust defense strategies to protect unmanned stores from adversarial threats. Highlighting the limitations of the current defense mechanisms in real-time detection systems and discussing various proactive measures, we provide insights into improving the robustness of object detection models and fortifying unmanned retail environments against these attacks.
\end{abstract}

}

\maketitle
\IEEEdisplaynontitleabstractindextext
\IEEEpeerreviewmaketitle

\section{Introduction}\label{sec1}
\IEEEPARstart{U}{nmanned} stores operate without sellers or employees, utilizing artificial intelligence-(AI-)based automation technologies. These stores provide consumers with a convenient shopping experience and effectively reduce operational costs, thus playing a crucial role in shaping the future of automated retail\cite{del2023phygital,wu2024service}. They leverage computer vision and deep neural networks (DNNs) to automate tasks such as product recognition, tracking, behavior analysis, and theft prevention. Enhanced detection capabilities are achieved using RGB and depth cameras, supplemented by sensors like RFID, LiDAR, and weight sensors \cite{bocanegra2020rfgo, duan2021full}. In 2018, Amazon launched Amazon Go, an unmanned store chain in the U.S., utilizing their Just Walk Out technology \cite{wankhede2018just}. However, high installation costs due to numerous sensors posed challenges \cite{amazongo}. To mitigate these costs, Amazon replaced depth cameras and load cell sensors with RGB cameras in 2023. Similarly, various global companies are adopting RGB cameras in unmanned stores to lower installation and operational expenses \cite{standardai}. While RGB data minimizes the need for additional information such as depth, weight, and vibration, fully unmanned stores still require DNN-based object detection technologies for automatic payment processing. However, DNNs are highly vulnerable to adversarial patch attacks, which are feasible in realistic settings like unmanned stores \cite{brown2017adversarial,wei2023unified}.

Adversarial patch attacks introduce noise into specific image areas to disrupt object detection models, posing significant security threats in automated retail. Attackers can print these patches and attach them to their bodies or expose them to cameras in unmanned stores \cite{lee2019physical, hu2021naturalistic}. While existing literature mainly focuses on physical threats and cost issues, research on AI security robustness is limited \cite{wang2021evaluate}.

Therefore, this study evaluates adversarial patch attacks in fully unmanned store environments, revealing new security vulnerabilities for DNN-based object detectors accessible to the public. We tested real-time models using printed adversarial patches on shelves equipped with RGB cameras in a simulated unmanned store. Our study explored three types of attacks and introduced a new histogram loss function to enhance attack performance based on the external appearance information of target objects. While existing studies \cite{shrestha2023towards,du2022physical} primarily focus on Hiding attacks, we also examined the practical effects of Altering attacks, which misidentify objects, and Creating attacks, which detect nonexistent objects. We incorporated the external appearance information of target objects, which an attacker can possess in a fully unmanned store environment, to improve the effectiveness of these attacks.

Finally, we assessed the realistic threat potential of adversarial patch attacks by conducting additional attacks in a black-box setting. Attackers may face constraints such as not knowing the color distribution of objects and backgrounds in physical settings or lacking access to a target object detection model, which we considered in our experiments with transfer and shadow attacks. The contributions of this study are as follows.

\begin{itemize}
  \item \textbf{Investigation of adversarial patch types}: We implemented three types of adversarial patch attacks---Hiding, Creating, and Altering attacks---to identify security vulnerabilities in object detection models. Particularly, we proposed a new loss function, the color histogram similarity loss, utilizing the knowledge possessed by the attacker regarding the appearance of a target object.
  \item \textbf{Physical attacks in an unmanned store testbed}: We constructed a testbed simulating a real unmanned store environment. We conducted experiments on two datasets, observing the effectiveness of adversarial patch attacks over consecutive frames. This allowed us to identify the existence of robust and non-robust classes and confirm that attacks can be activated under various environmental factors, including patch size, angle, and lighting.
  \item \textbf{Attacks in a black-box environment}: To emphasize the practicality of adversarial patch attacks, we imposed constraints on the knowledge of the attacker on data distribution and target model information. Particularly, we explored model transfer and shadow attacks, demonstrating that while attacks are limited without direct access to the model, the effectiveness of the attacks can be improved.
\end{itemize}

\section{Related Work}\label{sec2}
\subsection{Adversarial Patch Attacks}\label{subsec2_1}
DNNs traditionally exhibit vulnerabilities to adversarial examples \cite{shafahi2018adversarial,szegedy2013intriguing} owing to the excessive linearity in the hyperplanes that are formed during training\cite{goodfellow2014explaining}. Adversarial examples are primarily focused on inducing misclassifications by the model, which involves the injection of imperceptible perturbations into the original data, typically based on the $L_{p}$-norm\cite{carlini2017towards,wang2023adversarial}. Consequently, these examples are predominantly utilized to test the vulnerabilities of models or to enhance their robustness.

Conversely, adversarial patches are being extensively studied for deployment in physical environments. Similar to adversarial examples, these patches are intentionally designed to be noticeable to the human eye and are used to induce malfunctions in models. Recent studies \cite{zhu2023tpatch,hu2021naturalistic} have focused on enhancing the stealthiness of adversarial patches by refining loss functions to allow patches to assume more semantic or natural appearances. Moreover, the introduction of printable loss functions \cite{sharif2016accessorize} and specific conditions \cite{tsuruokawip} to ensure the effectiveness of attacks in physical settings presents realistic scenarios that reflect continual advancements in this field.

\subsection{Physical Adversarial Patch Attacks on Object Detection Models}\label{subsec2_2}
Adversarial patches pose a significant threat to object detection models. Liu et al. \cite{liu2018dpatch} found that adversarial patches \cite{brown2017adversarial} affecting classification models are ineffective against detection models, highlighting the need to target both bounding box regression and object classification. Initial physical adversarial patch attacks focused on reducing person detection accuracy by attaching patches to clothing. Thys et al. \cite{thys2019fooling} used a loss function targeting the objectiveness score and incorporated total variation (TV) loss to smooth patches and adjusted color values within the RGB spectrum for printability. Meanwhile, Adversarial patches have also disrupted unmanned aerial surveillance. Du et al. \cite{du2022physical} reported the first physical adversarial attacks in aerial scenes, examining the effects under various atmospheric conditions. Shrestha et al. \cite{shrestha2023towards} attached patches to unmanned aerial vehicle imagery, considering environmental noise and applying scene intensity adjustments and affine transformations during patch training. Tang et al. \cite{tang2024adversarial} introduced a patch-applier system for false-positive creation attacks in aerial scenes.

Despite these extensive studies, a gap remains in security analyses for automated retail environments like unmanned stores. Wang et al. \cite{wang2021universal} conducted comprehensive experiments on adversarial patch attacks in digital and physical settings, targeting Automatic Checkout scenarios. They extracted texture information from challenging samples and optimized patches to diminish structural similarity in attention maps, inspired by model attention bias. Unlike previous studies focusing on Hiding attacks tested on platforms like Taobao and JD.com, our research includes a broader analysis of Hiding, Creating, and Altering attacks. We observe the impact on real-time object detection models in testbed settings and propose a more realistic performance metric based on bounding box size rather than class classification performance.

\begin{figure*}[t!]
\centering
\includegraphics[width=0.85\textwidth]{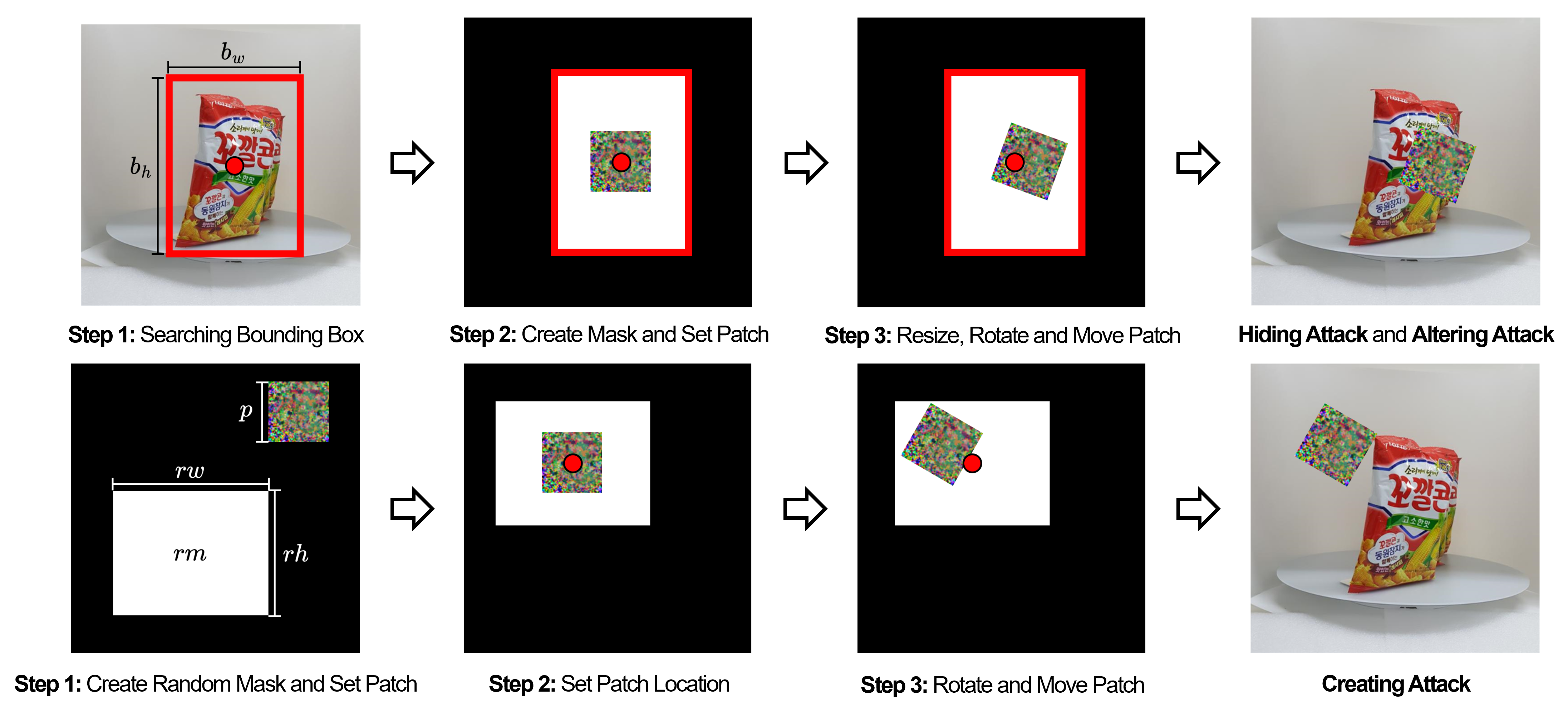}
\caption{Procedure of adversarial patch initialization.}
\label{Fig01}
\end{figure*}

\section{Methodology}\label{sec3}
\subsection{Preliminaries}\label{subsec3_1}
\subsubsection{Victim Object Detector}\label{subsubsec3_1_1}
Object detection models suitable for unmanned stores include one-stage detectors like the YOLO series and two-stage detectors like Faster region convolutional neural network (R-CNN) \cite{ren2015faster}. This study uses You only look once (YOLO) v5 \cite{jocher2021ultralytics}, which retains the Non-Maximum Suppression (NMS) \cite{hosang2017learning} process to reduce redundant bounding boxes by removing duplicates based on the objectiveness score, keeping only the box with the highest score. Specifically, an input image $x\in\mathbb{R}^{W\times H\times 3}$ is divided into an $s\times s$ grid to predict bounding boxes $B$, formatted as $(b_{x}, b_{y}, b_{w}, b_{h}, y_{obj})$. The NMS stage eliminates redundant boxes based on the Intersection over Union (IoU). Finally, the object class probability of each box is predicted as $y_{cls}=P(class_{c}|object)$, yielding $y=y_{obj}\times y_{cls}$.

Faster R-CNN operates in two phases: the Region Proposal Network (RPN) generates region proposals, and then classifies the object and refines its position. The RPN, integrated with a convolutional neural network (CNN), enhances detection performance and efficiency by embedding the proposal mechanism within the neural network. Specifically, the input image $x$ undergoes CNN and region of interest pooling to produce a resized feature map $R$. Class confidence $y_{cls}$ and object presence $y_{obj}$ are then detected, followed by NMS to finalize the output $y$. Both YOLO v5 and Faster R-CNN use $y_{cls}$ and $y_{obj}$, making them targets for adversarial patch attacks.

\subsubsection{Threat Model}\label{subsubsec3_1_2}
We defined a threat model by outlining the attackers' goals, knowledge, and capabilities to induce misbehaviors in the victim's object detection model using adversarial patches. The attackers may be in a white-box environment, where they can observe $y_{cls}$ and $y_{obj}$, or in a black-box environment, where they only have access to the outputs of a real-time object detection model. Additionally, we assumed that they have knowledge of the color and shape of the target class object, reflecting realistic scenarios in fully unmanned stores.

\subsubsection{Threat Scenario}\label{subsubsec3_1_3}
Adversarial patch attacks can significantly threaten fully unmanned stores, leading to theft, reputational damage, and financial losses. In a Hiding attack, a patch conceals high-value items, preventing detection during theft \textbf{(theft concealment)}, causing brand misidentification \textbf{(product counterfeiting)}, or interfering with product launches \textbf{(promotion and product launch interference)}. In a Creating attack, the patch makes the system recognize nonexistent products, leading to inventory errors \textbf{(inventory discrepancies)} and unintended discounts \textbf{(fake promotions)}. In an Altering attack, the patch cause misclassification and incorrect pricing, harming competitors' reputations \textbf{(brand damage)}, confusing customers \textbf{(customer confusion)}, and reducing customer satisfaction \textbf{(incorrect recommendations)}. This study explored how various threat scenarios can be realistically achieved by investigating adversarial patch attacks in both digital and physical environments.

\subsection{Adversarial Patch Initialization}\label{subsec3_2}
For a benign image $x$, an initial adversarial patch $p_{0}\in\mathbb{R}^{s\times s\times 3}$ is set, where $s$ is the patch size. We consider two initial settings for $p_{0}$: a gray patch (all pixel values set to 0.5) and a random patch (configured based on a Gaussian distribution). Additionally, a mask matrix $m\in {0,1}^{n}$ is used to attach the adversarial patch.
\begin{align}
m=f_{angle}\cdot f_{scale}\cdot f_{loc}(m_{0})
\end{align}\label{eq2}
The mask settings differ based on the attack type. For Hiding and Altering attacks, the patch is placed at a random location within the object's bounding box, as per the affine transformation approach described by Shrestha et al. \cite{shrestha2023towards}. In contrast, for Creating attacks, a comprehensive mask is required to activate both the object and background areas, as illustrated in Fig. \ref{Fig01}.

The goal of the Creating attack is to place one adversarial patch per image that can generate an incorrect bounding box. To achieve this, a random mask area $rm\in\mathbb{R}^{rw\times rh\times 1}$ is defined, where $rw$ and $rh$ are the width and height of the patch scaled by a random factor between $20$ and $70\%$ of image dimensions $W$ and $H$, respectively. Following this, the size of the patch is preset as follows:
\begin{align}
p_{size}=min(rw,rh)\times rand(0.25,0.40)
\end{align}\label{eq3}
In essence, the size of the patch is derived by scaling the smaller dimension of the mask area $rm$, which is either the width or the height, by a factor within the range of $[0.25, 0.4]$. Subsequently, the location of the patch is set based on the center of the mask area $rm$. To ensure that the patch remains within the mask area when rotated by a maximum angle of $45$ degrees via $f_{angle}$, the formula $l_{diag}=p_{size}\times (\sqrt{2}-1)$ is introduced to determine the center point of the patch. This calculation accounts for the diagonal length required to maintain the patch within the bounds during rotation:
\begin{align}
&p_{x}^{center}\in [\frac{1}{2}\times rw+l_{diag}, 1-\frac{1}{2}\times rw-l_{diag}]\times W\\,
&p_{y}^{center}\in [\frac{1}{2}\times rh+l_{diag}, 1-\frac{1}{2}\times rh-l_{diag}]\times H.
\end{align}\label{eq4}
Finally, using the center points $p_{x}^{center}$ and $p_{y}^{center}$ as references, the patch undergoes an affine transformation using the same $f_{loc}$ and $f_{angle}$ parameters as those used in the other types of attacks. This transformation effectively positions and orients the patch correctly, resulting in a final mask $m$.

To consider physical environmental factors during adversarial patch training, we applied random contrast and brightness transformations to each instance, and injected noise through a uniform distribution, following the scene intensity matching approach of Shrestha et al. \cite{shrestha2023towards}:
\begin{align}
p=p_{0}\times f_{const}+f_{bright}+f_{noise}.
\end{align}\label{eq5}
Finally, each instance $x_{adv}$ used for adversarial patch training was formed as follows:
\begin{align}\label{eq6}
x_{adv}=(1-m)\bigodot x+m\bigodot p,
\end{align}
where $\bigodot$ denotes the element-wise multiplication. The adversarially crafted instance $x_{adv}$ that was generated using (\ref{eq6}) was consistently used across all phases of patch training, validation, and testing. It was also employed to evaluate the attack performance in all digital environments that are described in Section \ref{subsec5_1}.

\subsection{Fundamental Components of Adversarial Patch Loss}\label{subsec3_3}
In this section, we define the loss functions required to train the three types of adversarial patches. Initially, the adversarial detection loss $L_{adv}$ used $y_{cls}$ and $y_{obj}$ to induce misbehaviors in each type of victim object detection model, which is elaborated in Section \ref{subsubsec3_4_1}. In addition, executing a physically printed adversarial patch attack is significantly more challenging than a digital attack. For a digital patch's effects to manifest physically, a printer must precisely reproduce all colors, and a camera must correctly identify all colors and noise in the patch. To address these challenges, we use loss terms $L_{TV}$ to minimize pixel differences and $L_{NPS}$ to maximize patch printability.

First, $L_{TV}$, which denotes the TV loss\cite{rudin1992nonlinear}, smoothens the texture of the patch to maintain overall consistency, while naturally blending sudden color changes or boundaries to obtain a more realistic appearance.
\begin{align}
L_{TV}=\frac{1}{i\times j}\Sigma_{i,j}((p_{i,j+1}-p_{i,j})^{2}+(p_{i+1,j}-p_{i,j})^{2})^{\frac{1}{2}}
\end{align}
Here, $p_{i,j}$ represents the value at the pixel $(i,j)$ in the adversarial patch.

$L_{NPS}$, introduced in \cite{sharif2016accessorize}, measures how well each patch color matches a list of printable colors $c_{print}$:
\begin{align}
L_{NPS}=\frac{1}{i\times j}\Sigma_{i,j}|p_{i,j}-c_{print}|
\end{align}

\subsection{Advanced Components of Adversarial Patch Loss}\label{subsec3_4}
\subsubsection{Adversarial Detection Loss Function}\label{subsubsec3_4_1}
Thys et al. \cite{thys2019fooling} and Du et al. \cite{du2022physical} found that optimizing $L_{adv}$ for Hiding attacks by minimizing $y_{obj}$ yields the most effective results. Conversely, Hu et al. \cite{hu2021naturalistic} and Shrestha et al. \cite{shrestha2023towards} induced misbehavior by simultaneously minimizing $y_{obj}$ and $y_{cls}$. To reconcile these differing findings, we evaluated the optimal use of $y_{obj}$ and $y_{cls}$ in the optimization process for adversarial patches. Consequently, the adversarial detection loss $L_{adv}$ for Hiding attacks was formulated as follows:
\begin{align}\label{formula12}
L_{adv}=\begin{cases}
max(y_{cls}) & \text{if use only}\;y_{cls} \\
max(y_{obj}) & \text{if use only}\;y_{obj} \\
max(y_{obj}\cdot max(y_{cls})) & \text{if use both}\;y_{cls}, \;y_{obj}.
\end{cases}
\end{align}
Essentially, the optimization process for each object within an image involves observing either the highest class confidence $max(y_{cls})$ or highest objectiveness score $max(y_{obj})$ in the area where the patch is applied. Particularly, the goal of the Hiding attack, which is to ensure that the object is not detected as any class, is pursued by measuring these maxima in each iteration and driving all values toward minimization.

\begin{figure}[t!]
\centering
\includegraphics[width=0.48\textwidth]{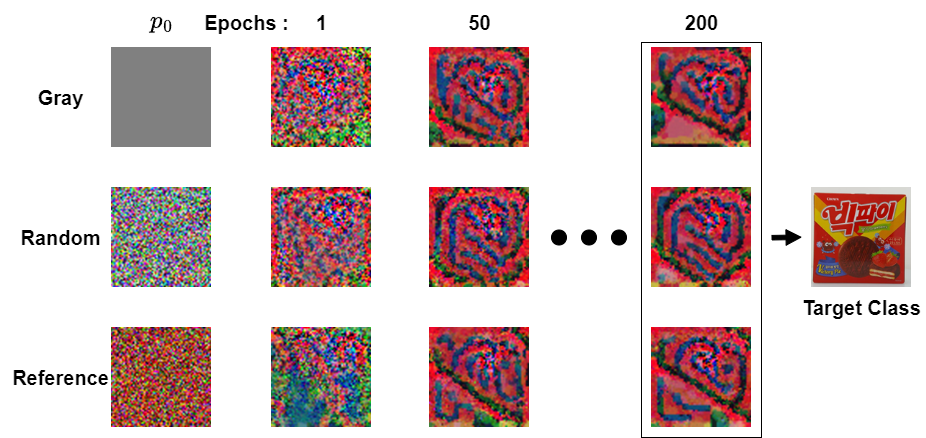}
\caption{Adversarial patch generation process per $p_{0}$ according to the epochs. All generated patches are similar in color to the target class object.}
\label{Fig03}
\end{figure}

\begin{table}[t!]
\centering
\caption{Average attack convergence rate $R_{C}$ and attack success rate $R_{S}$ per $p_{0}$}
\label{Table01}
\begin{tabular}{@{}cccc@{}}
\toprule
\textbf{$p_{0}$} & \textbf{gray} & \textbf{random} & \textbf{reference} \\ \midrule
\textbf{$R_{C}$} & 140.191       & 126.619         & 85.714             \\ \midrule
\textbf{$R_{S}$} & 0.2550        & 0.2576          & 0.2428             \\ \bottomrule
\end{tabular}
\end{table}

As defined in Section \ref{subsec3_2}, the Creating attack aims to induce detection of objects in arbitrary locations within an image, regardless of the original class. Consequently, both $y_{obj}$ and $y_{cls}$ must be maximized simultaneously for this type of attack. Therefore, the adversarial detection loss $L_{adv}$ for the Creating attack is formulated as follows:
\begin{align}
L_{adv}=1-max(y_{obj}\cdot y^{t}_{cls}),
\end{align}
where $y^{t}_{cls}$ denotes the confidence score of the target class.

Finally, the Altering attack aims to maximize the confidence score $y_{cls}^{t}$ for the target class while minimizing the confidence scores for all other classes, thereby ensuring that only the target class is detected. As the bounding box of the target object must remain unchanged, the adversarial detection loss $L_{adv}$ for the Altering attack is defined by controlling only $y_{cls}$ as follows:
\begin{align}
L_{adv}=max(y_{cls}|cls\neq t)+2\cdot (1-y^{t}_{cls}).
\end{align}
Through an empirical evaluation, we discovered that starting the confidence score of the target class from zero and exceeding a certain threshold were challenging processes. Consequently, we decided to double the weight of the second term to ensure that it received a higher emphasis.

\subsubsection{Evaluating the Impact of Color Similarity on Adversarial Patch Effectiveness}\label{subsubsec3_4_2}
When attackers in environments such as unmanned stores know the color and shape of a target class, this knowledge can pose a potential threat in adversarial patch attacks. We observed through existing literature that known adversarial patches tend to converge to resemble the color or shape of the target class, such as zebras\cite{shrestha2023towards}, airplanes\cite{tang2024adversarial}, and birds\cite{zhang2023capatch}.

We analyzed the success rate of Altering attack patches using a YOLO v5 victim model trained on the snack dataset. We generated adversarial patches for 21 target classes with three initial values (gray, random, and reference), totaling 63 samples. For reference patches, we assumed the attacker knows the target class information and initialized $p_{0}$ using a Gaussian distribution with a standard deviation of 0.3 based on the target object's average RGB values.

Fig. \ref{Fig03} and Table \ref{Table01} show the epoch-by-epoch changes in patches and the average attack convergence rate ($R_{C}$) and success rate ($R_{S}$) for 21 target classes. During the last 10 epochs of each adversarial patch creation process, the maximum and minimum values of $R_{S}$ were extracted. This range was used to sequentially check whether $R_{S}$ from the first 10 epochs was within this range, designating the starting epoch of the captured moment as $R_{C}$. We observed that all patches converged to similar colors and shapes. The reference patch converged fastest, and the gray patch slowest, suggesting that knowledge of the target object's color allows for lower-cost adversarial patch creation.

\begin{figure}[t!]
\centering
\includegraphics[width=0.48\textwidth]{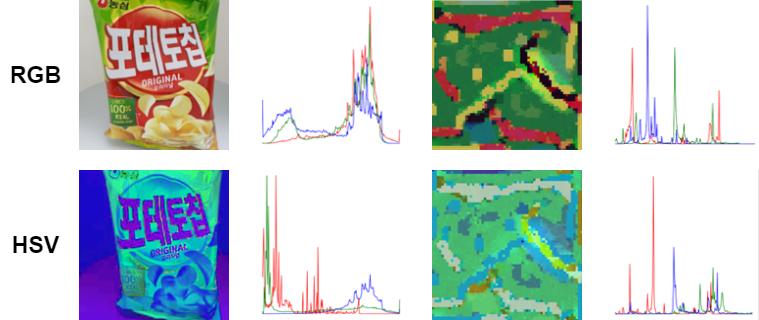}
\caption{Histograms of reference image and adversarial patch in RGB (top) and hue saturation value (bottom) spaces.}
\label{Fig04}
\end{figure}

\begin{table}
\centering
\caption{Correlation analysis between attack success rate and each similarity metric in RGB and hue saturation value spaces}
\label{Table02}
\begin{tabular}{@{}cccccc@{}}
\toprule
RGB                               & Corr   & Inter  & Bhattac  & Chi-S          & KLD    \\ \midrule
All $R_{S}$                    & -0.077 & -0.097 & -0.121   & -0.043         & -0.303 \\ \midrule
$0.1\leq R_{S}$ & 0.158  & -0.447 & -0.420   & 0.294          & -0.339 \\ \midrule
\textbf{HSV}                      & Corr   & Inter  & Bhattack & \textbf{Chi-S} & KLD    \\ \midrule
All $R_{S}$                    & -0.074 & -0.060 & -0.030   & \textbf{0.284} & -0.197 \\ \midrule
$0.1\leq R_{S}$ & -0.104 & -0.470 & -0.418   & \textbf{0.401} & -0.367 \\ \bottomrule
\end{tabular}
\end{table}

We then examined the relationship between the color similarity of patches and attack performance. Histograms in RGB and HSV color spaces were extracted from both target class images and adversarial patches (Fig. \ref{Fig04}). The similarity $S$ between histograms was measured using correlation (Corr), intersection (Inter), Bhattacharyya distance (Bhattac), chi-square (Chi-S), and Kullback--Leibler divergence (KLD) functions \cite{marin2016comparative}. If the actual similarity and numerical values were inversely proportional, the similarity was converted into $-S$. The correlation between similarity and attack success rate was then assessed to determine whether color similarity enhances attack performance.

Table \ref{Table02} lists the correlation results between attack success rate and each similarity measurement function. Chi-S-based similarity in HSV space showed the highest positive correlation (0.284), prompting further analysis for patches with $R_{S} >$ 0.1, yielding a significant correlation of 0.401. These findings suggest that incorporating a similarity-based loss function during adversarial patch training can enhance attack performance when the attacker knows the color of the target class object.

\begin{algorithm}[t!]
\centering
\caption{Color histogram similarity loss}\label{Alg1}
\begin{algorithmic}[1]
    \Require Adversarial Patch $p$
    \Ensure{CropObject}{$(x,y)$}: Crop an input image $x$ within the bounding boxes $y$.
    \Ensure{RGB2HSV}{$(x)$}: Convert RGB colors of $x$ to HSV.
    \Ensure{Histogram}{$(x,bins)$}: Calculate the histogram of the image $x$ with histogram bin boundaries $bins$.
    \renewcommand{\algorithmicensure}{\textbf{Output:}}
    \Ensure{$L_{His}$}
    \Statex
    \renewcommand{\algorithmicensure}{\textbf{Initialization:}}
    \Ensure{}
    \Statex $x_{ref}$ : Reference image
    \Statex $y_{ref}$ : The bounding boxes of reference image
    \State $\widehat{x}_{ref}$\;=\;CropObject($x_{ref},y_{ref}$)
    \State $\overline{x}_{ref}$\;=\;RGB2HSV($\widehat{x}_{ref}$)
    \State $\overline{H}_{ref}$\;=\;Histogram($\overline{x}_{ref},\, [0,256]$)\\
    \Return $\overline{H}_{ref}$
    \Statex
    \renewcommand{\algorithmicensure}{\textbf{Run:}}
    \Ensure{}
    \State $\overline{p}$\;=\;RGB2HSV($p$)
    \State $\overline{H}_{p}$\;=\;Histogram($\overline{p},\, [0,256]$)
    \State $L_{His}=d(\overline{H}_{ref},\,\overline{H}_{p})\,=\,\Sigma_{I}\frac{(\overline{H}_{ref}(I)-\overline{H}_{p}(I))^{2}}{\overline{H}_{ref}(I)+\overline{H}_{p}(I)}$\\
    \Return $L_{His}$
\end{algorithmic}
\end{algorithm}

\subsubsection{Color Histogram Similarity Loss Function}\label{subsubsec3_4_3}
Therefore, we propose a new loss function $L_{His}$ that updates adversarial patches using the Chi-S-based color histogram similarity in the HSV space between the target object and adversarial patch. The algorithm is presented in Algorithm \ref{Alg1}. Initially, during the initialization stage, the reference image of the target class object $x_{ref}$ and its bounding box $y_{ref}$ are used to crop the object area. The cropped image is then converted to HSV space, from which a histogram $\overline{H}_{ref}$ consisting of 256 columns is obtained. In the run stage, the histogram $\overline{H}_{p}$ of the input adversarial patch $p$ is obtained using the same process; then, the alternative Chi-S statistic $L_{His}$ is measured between the two histograms to minimize it, which is used to update the patch. This loss function is applied during the implementation of Creating and Altering attacks to induce detection as a target class.

\subsubsection{Implementation of the Final Adversarial Patch Loss Function}\label{subsubsec3_4_4}
The final adversarial patch loss function was calculated based on the weight $\lambda$ of each loss term, which was determined through an empirical evaluation to achieve optimal values. For the Hiding attack, the function is defined as
\begin{equation}
\begin{split}
    L_{Hiding}=&\lambda_{adv}\cdot L_{adv}+\lambda_{sal}\cdot L_{sal}\\
    &+\lambda_{TV}\cdot L_{TV}+\lambda_{NPS}\cdot L_{NPS},
\end{split}
\end{equation}
where $\lambda_{adv}$, $\lambda_{sal}$, $\lambda_{TV}$, and $\lambda_{NPS}$ were set to 3.0, 1.0, 1.0, and 1.0, respectively. Furthermore, in Section \ref{subsec5_4_2}, we confirm that simultaneously optimizing $y_{cls}$ and $y_{obj}$ in $L_{adv}$ resulted in the highest attack success rate. Consequently, the experimental results presented in Section \ref{sec5} reflect the outcomes for $L_{adv}=\max(y_{obj} \cdot \max(y_{cls}))$.

For the Creating attack, the loss function is defined as
\begin{equation}
\begin{split}
    L_{Creating}=&\lambda_{adv}\cdot L_{adv}+\lambda_{TV}\cdot L_{TV}\\
    &+\lambda_{NPS}\cdot L_{NPS}+\lambda_{His}\cdot L_{His},
\end{split}
\end{equation}
where $\lambda_{adv}$, $\lambda_{TV}$, $\lambda_{NPS}$, and $\lambda_{His}$ were set to 3.0, 0.5, 1.0, and 0.3, respectively.

Finally, the configuration used in $L_{Creating}$ was used in the Altering attack with identical weights for each loss term.
\begin{equation}
\begin{split}
    L_{Altering}=&\lambda_{adv}\cdot L_{adv}+\lambda_{TV}\cdot L_{TV}\\
    &+\lambda_{NPS}\cdot L_{NPS}+\lambda_{His}\cdot L_{His}
\end{split}
\end{equation}

\section{Experimental Setup}\label{sec4}
\subsection{Datasets}\label{subsec4_1}
In this study, we implemented a physical testbed and selected a detection system for snacks and fruits, typical items sold in fully unmanned stores. We gathered the data using Roboflow\cite{roboflow}, a tool for computer vision tasks. The snack dataset included 21 categories of popular snacks, chosen for their distinct colors and shapes, suitable for real-world multiple object detection tasks. This dataset was divided into 1,676 training sets, 479 validation sets, and 239 test sets. The training set was subjected to four types of data augmentation processes: horizontal flipping, brightness alteration, affine rotation, and coarse dropout, effectively quadrupling the dataset.

Similarly, the fruit dataset, comprising 10 categories of common fruits, was divided into 1,550 training sets, 430 validation sets, and 237 test sets, with the same augmentation processes applied. Both datasets were resized to $1088\times 1088$ pixels in RGB color channels, ensuring robust training data for accurate object recognition and classification under varied conditions, reflecting practical retail environments.

\subsection{Victim Object Detector}\label{subsec4_2}
We selected YOLO v5l6 (YOLO v5 (L)) and Faster R-CNN with 50-layer residual network\cite{he2016deep} as the backbone for our victim object detectors, and both models were trained on the snack and fruit datasets. Furthermore, for the shadow attack discussed in Section \ref{subsec5_3_2}, we set YOLO v5x6 (YOLO v5 (X)) and Faster R-CNN with MobileNet v3\cite{howard2019searching} as the backbone in the shadow models. These shadow models were trained on the snack dataset using the outputs of victim object detectors as the ground truth to mimic their latent distribution. The YOLO models used in our experiments were all retrained using the pre-trained models provided by Ultralytics version 8.0, whereas the Faster R-CNN models were retrained using the pre-trained models available in torchvision version 0.17.

\begin{table*}[t!]
\centering
\caption{Effectiveness of Adversarial Patches based on Attack Type for Each Dataset and Model in a Digital Environment}
\label{Table03}
\resizebox{\textwidth}{!}{%
\begin{tabular}{@{}ccccccccccccccc@{}}
\toprule
                       &                               &              & \multicolumn{4}{c}{Hiding Attack} & \multicolumn{4}{c}{Creating Attack} & \multicolumn{4}{c}{Altering Attack} \\ \midrule
Dataset                & Model                         & Attack       & $mAP$  & $mAR$  & $CM$   & $CIoU$ & $mAP$   & $mAR$  & $CM$   & $CIoU$  & $mAP$   & $mAR$  & $CM$   & $CIoU$  \\ \midrule
\multirow{8}{*}{Snack} & \multirow{4}{*}{Yolo v5}      & No Attack    & 0.996  & 0.998  & -      & -      & 0.996   & 0.998  & -      & -       & 0.996   & 0.998  & -      & -       \\ \cmidrule(l){3-15} 
                       &                               & Noise & 0.659  & 0.682  & 0.336  & 0.272  & 0.856   & 0.873  & 0.003  & 0.020   & 0.659   & 0.682  & 0.001  & 0.016   \\ \cmidrule(l){3-15} 
                       &                               & Gray Patch   & 0.079  & 0.083  & 0.911  & 0.906  & 0.570   & 0.604  & 0.347  & 0.196   & 0.349   & 0.387  & 0.176  & 0.255   \\ \cmidrule(l){3-15} 
                       &                               & Random Patch & 0.111  & 0.121  & 0.906  & 0.842  & 0.559   & 0.597  & 0.333  & 0.192   & 0.348   & 0.383  & 0.180  & 0.252   \\ \cmidrule(l){2-15} 
                       & \multirow{4}{*}{Faster R-CNN} & No Attack    & 0.996  & 0.998  & -      & -      & 0.996   & 0.998  & -      & -       & 0.996   & 0.998  & -      & -       \\ \cmidrule(l){3-15} 
                       &                               & Noise & 0.456  & 0.516  & 0.657  & 0.291  & 0.733   & 0.764  & 0.023  & 0.024   & 0.456   & 0.516  & 0.033  & 0.068   \\ \cmidrule(l){3-15} 
                       &                               & Gray Patch   & 0.367  & 0.415  & 0.737  & 0.404  & 0.644   & 0.681  & 0.204  & 0.181   & 0.318   & 0.368  & 0.133  & 0.160   \\ \cmidrule(l){3-15} 
                       &                               & Random Patch & 0.331  & 0.376  & 0.815  & 0.435  & 0.632   & 0.669  & 0.229  & 0.200   & 0.323   & 0.373  & 0.136  & 0.171   \\ \midrule
\multirow{8}{*}{Fruit} & \multirow{4}{*}{Yolo v5}      & No Attack    & 0.994  & 0.996  & -      & -      & 0.994   & 0.996  & -      & -       & 0.994   & 0.996  & -      & -       \\ \cmidrule(l){3-15} 
                       &                               & Noise & 0.603  & 0.636  & 0.394  & 0.416  & 0.844   & 0.867  & 0.006  & 0.022   & 0.603   & 0.636  & 0.008  & 0.062   \\ \cmidrule(l){3-15} 
                       &                               & Gray Patch   & 0.065  & 0.076  & 0.951  & 0.866  & 0.621   & 0.688  & 0.446  & 0.154   & 0.239   & 0.289  & 0.234  & 0.154   \\ \cmidrule(l){3-15} 
                       &                               & Random Patch & 0.081  & 0.085  & 0.906  & 0.926  & 0.611   & 0.683  & 0.479  & 0.127   & 0.254   & 0.307  & 0.212  & 0.120   \\ \cmidrule(l){2-15} 
                       & \multirow{4}{*}{Faster R-CNN} & No Attack    & 0.994  & 0.981  & -      & -      & 0.994   & 0.981  & -      & -       & 0.994   & 0.981  & -      & -       \\ \cmidrule(l){3-15} 
                       &                               & Noise & 0.267  & 0.328  & 0.704  & 0.437  & 0.693   & 0.752  & 0.020  & 0.021   & 0.267   & 0.328  & 0.024  & 0.062   \\ \cmidrule(l){3-15} 
                       &                               & Gray Patch   & 0.106  & 0.134  & 0.922  & 0.685  & 0.560   & 0.665  & 0.473  & 0.225   & 0.105   & 0.131  & 0.021  & 0.041   \\ \cmidrule(l){3-15} 
                       &                               & Random Patch & 0.135  & 0.164  & 0.908  & 0.666  & 0.557   & 0.662  & 0.424  & 0.192   & 0.104   & 0.131  & 0.020  & 0.043   \\ \bottomrule
\end{tabular}}
\end{table*}

\subsection{Physical Testbed of an Unmanned Store}\label{subsec4_3}
We constructed a testbed that closely simulates a real-world environment. We set up a suitably sized shelf capable of holding multiple objects and mounted a real-time surveillance webcam (ABKO APC930 QHD Webcam) that supports a maximum resolution of $2592 \times 1944$ and a video frame rate of 30 fps. Moreover, we placed actual products that were listed in the snack dataset on the shelf while using realistic models of fruits that were shown in the fruit dataset.

The various environmental issues that may arise when conducting adversarial patch attacks in a physical setting must be considered. To address this, we compared attacks under both dark and bright lighting conditions using artificial lighting and recorded the performance of the patch when captured from the front and side angles. Moreover, we printed patches in sizes of $10 \times 10$ and $5 \times 5$ to verify the performance differences based on the perceived size of the patches.

\subsection{Attack Measurements}\label{subsec4_4}
We analyzed the effectiveness of adversarial patches using four evaluation metrics. First, we measured the mean average precision ($mAP$) and mean average recall ($mAR$), which are widely used metrics for object detection models, based on IoU thresholds ranging from 0.50 to 0.95. However, these two metrics alone exhibit limitations when measuring the attack performances of the three types of adversarial patches. Consequently, we introduced two new evaluation metrics. First, we captured the changes in the object class detection for each attack type based on a $CM$, and quantified them using the true-positive rate ($TPR$) and false-positive rate ($FPR$) according to the number of objects in the test image $N_{O}$. For the Hiding attack, the disappearance rate of the objects can be measured by $CM=1-(TPR/N_{O})$. For the Creating attack, the $FPR$ of previously undetected target class objects can be measured by $CM=FPR/N_{O}$. Moreover, for the Altering attack, the rate at which the original object disappears while the target class object is detected can be measured by $CM=(1-(TPR/N_{O}))\cap (FPR/N_{O})$.

Furthermore, we introduce a metric, complete IoU ($CIoU$), to judge the naturalness of the bounding boxes generated by adversarial patches. Even if an attack successfully induces false detections, if the size of the bounding box is absurdly small as compared to that of the target object, such attacks are likely to be detected by anomaly detection systems. Therefore, we measured the similarity between the sizes of the generated bounding box $B_{P}$ and original bounding box $B_{GT}$ using CIoU \cite{zheng2020distance}. The $CIoU$ for each patch type is measured as follows. For the Hiding attack, $CIoU =1-(B_{GT}\cap B_{P})$, which measures the decrease in the bounding box from the original. For the Creating attack, the random mask area $rm$, which is described in Section \ref{subsec3_2}, was assumed to be $B_{GT}$, and $CIoU=B_{GT}\cap B_{P}$ was measured. For the Altering attack, the bounding box should be preserved while only inducing misclassification of the target class. Therefore, we evaluated the extent to which the size of the original bounding box was preserved using $CIoU=B_{GT}\cap B_{P}$.

\subsection{Implementation Details}\label{subsec4_5}
All experiments were conducted on an NVIDIA GeForce RTX 3090 GPU using Python 3.9 with PyTorch 2.1.0 and torchvision 0.16.0. Particularly, to implement adversarial patch updates, we set the initial patch size to $s=64$. To optimize the loss, we employed the Adam optimizer with a starting learning rate of 0.03 and a ReduceLROnPlateau scheduler with a patience number of 50, iterating for a total of 200 epochs.

\section{Experimental Results}\label{sec5}
\subsection{Adversarial Patch Attacks in a Digital Environment}\label{subsec5_1}
\subsubsection{Hiding Attack}\label{subsec5_1_1}
Table \ref{Table03} lists the performance measurements of the three types of adversarial patches in each environment. The results indicate that the Hiding attack is effective in all cases. YOLO v5 shows robustness against random noise-based occlusions but is highly vulnerable to adversarial patches. Conversely, Faster R-CNN exhibits significant performance degradation with random noise, and the use of adversarial patches exacerbates this vulnerability. However, Faster R-CNN maintains a relatively robust $CIoU$ across both datasets, indicating its ability to effectively preserve bounding boxes by locally optimizing candidate anchor boxes based on feature extraction through the RPN, unlike YOLO, which uses grid-based searches.

\begin{figure}[t!]
\centering
\includegraphics[width=0.4\textwidth]{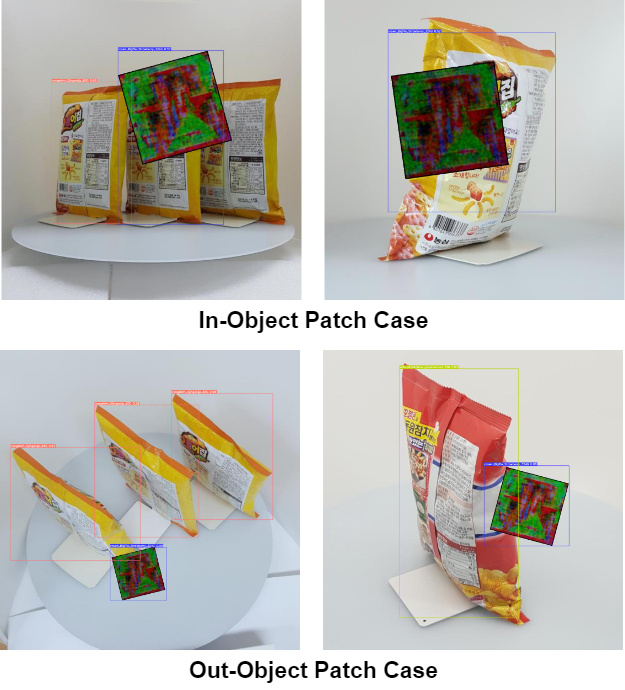}
\caption{Bounding box generation results of adversarial patches for Creating attacks in a digital environment based on patch location. Blue bounding boxes indicate the target class, whereas other colors represent the original classes.}
\label{Fig06}
\end{figure}

\begin{figure*}[t!]
\centering
\includegraphics[width=1.0\textwidth]{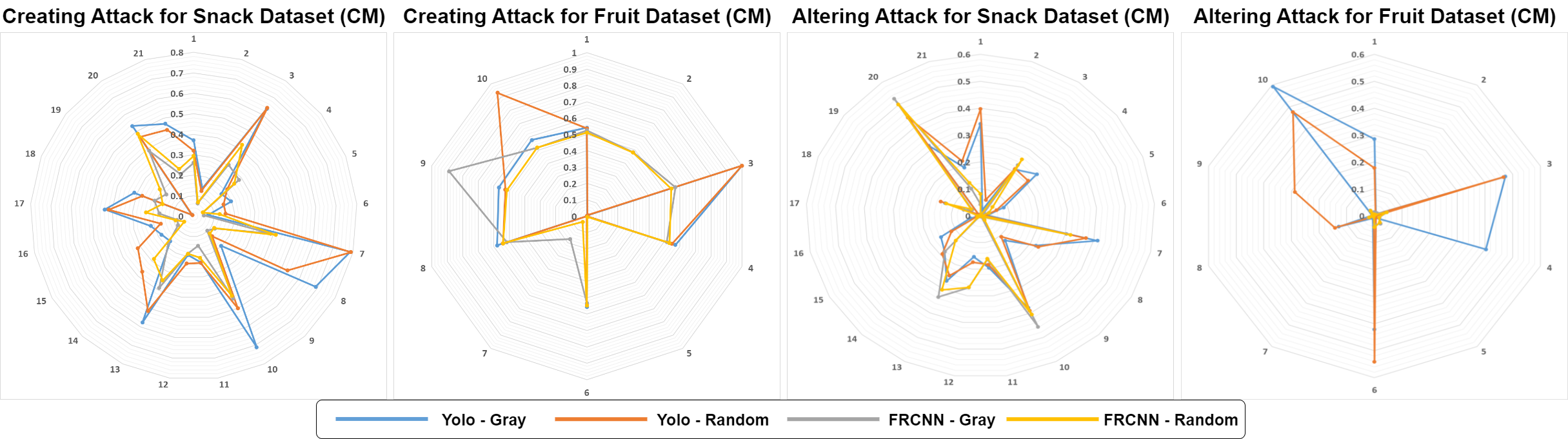}
\caption{Visualization of performance differences in adversarial patch attacks based on target classes.}
\label{Fig07}
\end{figure*}

\subsubsection{Creating Attack}\label{subsec5_1_2}
Conversely, the Creating attack shows effective results in all cases. Notably, because only one patch is generated per image during the Creating attack, the decrease in $mAP$ and $mAR$ for both random noise and the patch is lower compared to other attack types. Moreover, the nonzero $CM$ and $CIoU$ values for random noise reflect the inherent false positives of the victim model. However, the $CM$ and $CIoU$ for the patch show increments of up to 0.453 and 0.204, respectively, over random noise in the Faster R-CNN fruit dataset. These increments are sufficiently significant to undermine the availability of real-time monitoring systems in fully unmanned stores. Moreover, the $CIoU$ levels around 0.2 across all cases indicate the limitation of the creating patch in generating bounding boxes close to the actual object size. Particularly, Fig. \ref{Fig06} shows the cases in which the patch overlaps with the object (In-Object Patch Case) and is located in the background (Out-Object Patch Case). The blue bounding boxes represent the target classes, whereas the other colors represent the original classes. Here, bounding boxes similar in size to the actual object are effectively generated in the In-Object Patch Case, whereas the bounding boxes were constrained by the patch size in the Out-Object Patch Case. This insight suggests that defenders can detect anomalies based on the bounding box size; conversely, attackers are presented with a new challenge of increasing the bounding box size in Out-Object Patch Cases.

\subsubsection{Altering Attack}\label{subsec5_1_3}
The Altering attack is significantly more difficult than the other two types of attacks. This is because it can only control $y_{cls}$ to preserve the bounding box, while simultaneously minimizing the confidence values of the original and target classes, which were initially close to zero. As shown in Table \ref{Table03}, the attack succeeded in increasing $CM$ and $CIoU$ in the case of random noise in most cases; however, it completely failed in the Faster R-CNN fruit data case. We can observe that the $mAP$ and $mAR$ values decrease for random noise, indicating that the adversarial patch succeeded in causing missed detections but struggled to increase the confidence of the target class. Conversely, $CM$ increases by up to 0.226 in the YOLO v5 fruit data case, suggesting that this attack could pose a valid threat in specific scenarios depending on the threat context.

\subsubsection{Analysis of Attack Performance by Target Class}\label{subsec5_1_4}
In Table \ref{Table03}, the performances of the Creating and Altering attacks appear to be low. We identified that performance variation according to the target class is a major cause for this inferior performance and accordingly analyzed the differences in attack performance based on the class. Fig. \ref{Fig07} shows the class-specific $CM$ for each victim model and dataset for both the Creating and Altering attacks. The radial graphs for each case exhibit similar trends according to the class. The Creating attack achieves $CM$ values of up to 0.79 (random patch for $t=6$ on the YOLO model) and 1.00 (two patches for target $t=2$ on the YOLO model) in the snack and fruit datasets, respectively. Moreover, the Altering attack achieves $CM$ values of up to 0.526 (gray patch for target $t=19$ on the Faster R-CNN model) and 0.593 (gray patch for target $t=9$ on the YOLO model), implying that it can pose a significant threat to specific targets in a fully unmanned store. Furthermore, the similar vulnerabilities of the target classes within the same dataset, regardless of the object detection model structure, indicate an open challenge for attackers and defenders in identifying vulnerable classes or exploring mitigation strategies, such as class-agnostic patches, respectively.

\subsection{Adversarial Patch Attacks in a Physical Environment}\label{subsec5_2}
We empirically evaluated the threat posed by adversarial patches in a fully unmanned store testbed and observed their printability. Using the YOLO v5 model, we tested the detection of snacks and fruits. Each attack type was individually evaluated using five randomly selected original objects placed on shelves. Additionally, we assessed five arbitrary target classes for the Creating and Altering attacks. In each experimental case, we measured the average attack success rate by exposing the adversarial patches on the screen for 150 consecutive frames and calculating the corresponding $CM$.

\begin{figure}[t!]
\centering
\includegraphics[width=0.48\textwidth]{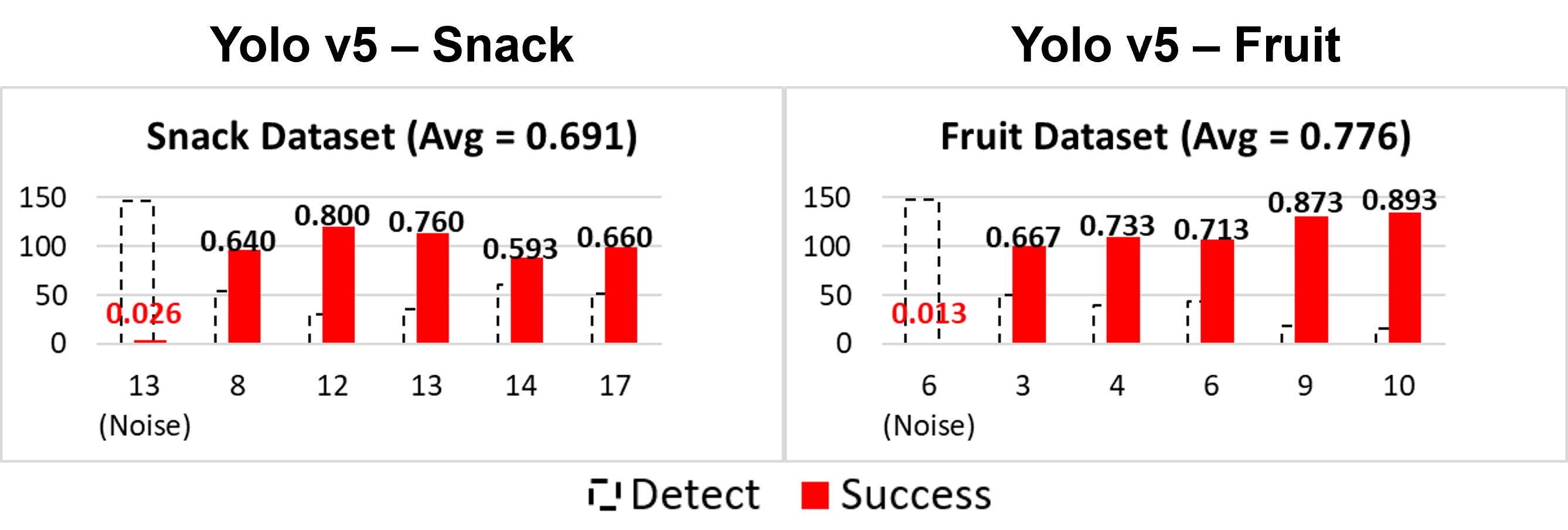}
\caption{Experimental results of the Hiding attack in a physical unmanned store testbed. The x-axis of each graph represents the original object, black dotted bars represent ``Detect,'' red bars represent ``Success,'' and numbers on each bar indicate the attack success rate ``$CM$'' for all frames.}
\label{Fig09}
\end{figure}

\begin{figure*}[t!]
\centering
\includegraphics[width=0.95\textwidth]{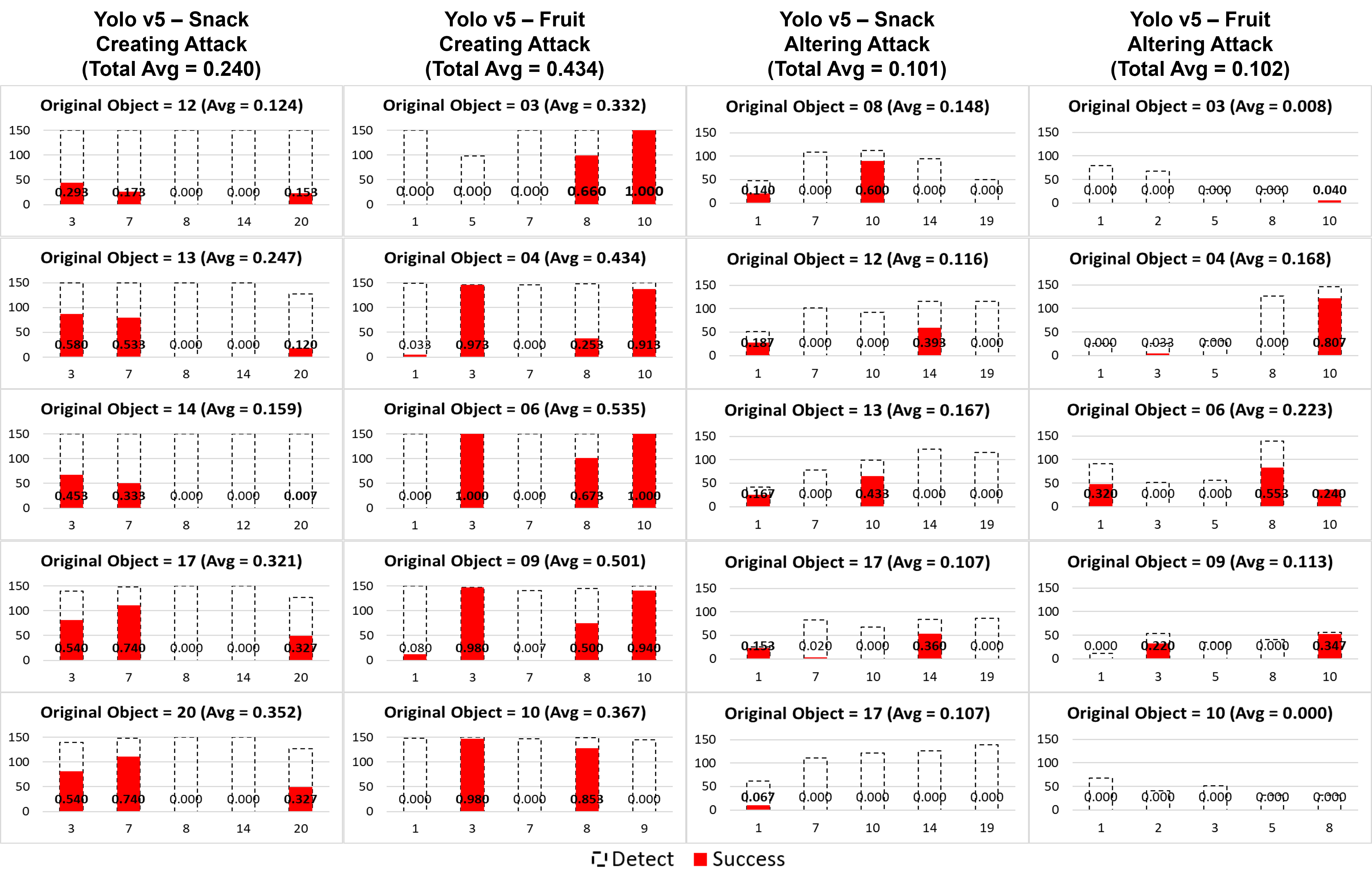}
\caption{Experimental results of Creating and Altering attacks in a physical unmanned store testbed. The x-axis of each graph represents the target class, black dotted bars represent ``Detect,'' red bars represent ``Success,'' and numbers on each bar indicate the attack success rate ``$CM$'' for all frames.}
\label{Fig10}
\end{figure*}

Fig. \ref{Fig09} shows the results of the Hiding attack for each original object. The x-axis represents the class number of the original object and the y-axis shows the values for all 150 frames. Here, the black dotted bar indicates the number of frames in which any object was detected, and the red bar indicates the number of frames in which the attack succeeded. The numbers on each bar represent the ``CM'' for all frames. The leftmost bars in each figure show the values when patches created from random noise were placed on the 13th class in the snack dataset and 6th class in the fruit dataset, which were correctly classified as the original class, except for four and two frames, respectively. This indicates that the victim model that we aimed to attack had relatively robust real-time detection capabilities and was resistance to occlusions. Adversarial patches showed average $CM$ of 0.691 and 0.776 for the two datasets, respectively, indicating that they could hide the presence of the object in over 60\% of the frames, despite class variance.

Fig. \ref{Fig10} shows the results of the Creating and Altering attack experiments for both datasets, categorized by each original object and target class. The x-axis represents the class number of the target class, and the number at the top of each graph shows the class number of the original object and its average $CM$. We can observe that the Creating attack did not succeed for certain target classes. However, certain classes (classes 3, 7, and 20 in the snack dataset and classes 3, 8, and 10 in the fruit dataset) exhibit high attack performances for most original objects, which were also the target classes with significant attack success rates, as shown in Fig. \ref{Fig07}. These results emphasize that attackers must pre-identify whether their desired target class is vulnerable or robust, and defenders must identify ways to improve the robustness of these vulnerable classes and ensure that they are not exposed externally.

Meanwhile, the Altering attack was effective for specific classes (1, 10, and 14 in the snack dataset and 1, 3, 8, and 10 in the fruit dataset). Additionally, we previously emphasized that to succeed in an Altering attack, the original class of the object must first be undetected, followed by misclassification to the target class, making this attack more challenging. The black dotted "Detect" in each graph shows a significant reduction in most cases, indicating that while the Altering attack may have failed, a certain degree of hiding effect was achieved.

\begin{figure*}[t!]
\centering
\includegraphics[width=1.0\textwidth]{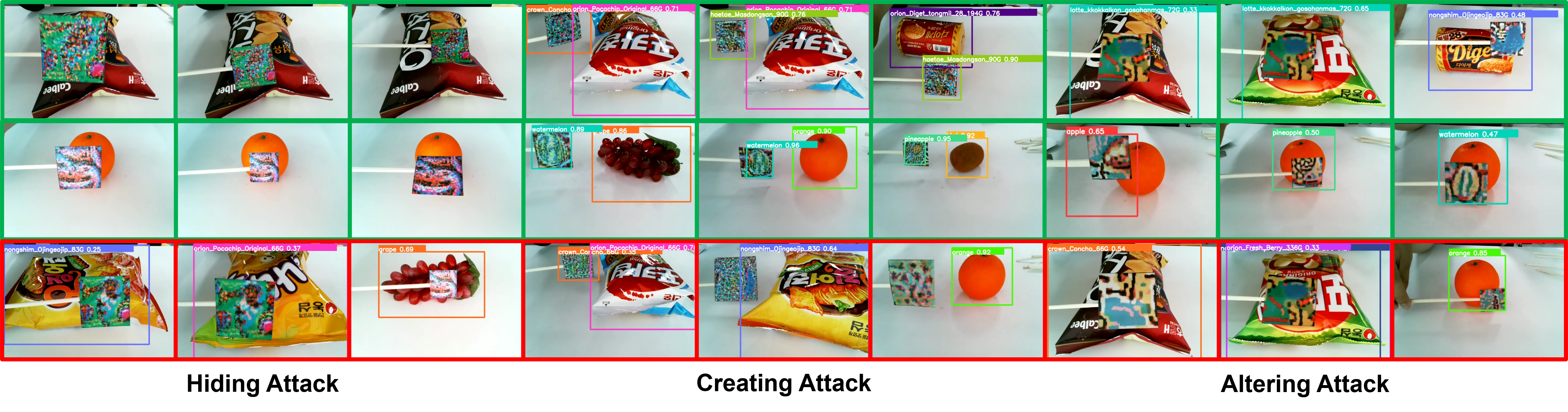}
\caption{Adversarial patch attacks on various original objects and target classes in a physical unmanned store testbed. Green borders indicate successful attacks, whereas red borders indicate failed attacks.}
\label{Fig11}
\end{figure*}

\begin{table*}[t!]
\centering
\caption{Model Transfer Attacks of Adversarial Patches in a Black-box Scenario}
\label{Table05}
\renewcommand{\arraystretch}{1.5}
\resizebox{\textwidth}{!}{%
\begin{tabular}{ccccccccccccccc}
\hline
                       &                                                                                            &                      & \multicolumn{4}{c}{Hiding Attack} & \multicolumn{4}{c}{Creating Attack} & \multicolumn{4}{c}{Altering Attack} \\ \hline
Dataset                & Model                                                                                      & Attack               & $mAP$  & $mAR$  & $CM$   & $CIoU$ & $mAP$   & $mAR$  & $CM$   & $CIoU$  & $mAP$   & $mAR$  & $CM$   & $CIoU$  \\ \hline
\multirow{4}{*}{Snack} & \multirow{4}{*}{\begin{tabular}[c]{@{}c@{}}Faster R-CNN \\ $\mapsto$ YOLO v5\end{tabular}} & Random Noise         & 0.659  & 0.682  & 0.336  & 0.272  & 0.856   & 0.873  & 0.003  & 0.020   & 0.659   & 0.682  & 0.001  & 0.016   \\ \cline{3-15} 
                       &                                                                                            & Gray Patch           & 0.571  & 0.595  & 0.497  & 0.307  & 0.767   & 0.762  & 0.009  & 0.018   & 0.534   & 0.560  & 0.004  & 0.037   \\ \cline{3-15} 
                       &                                                                                            & Random Patch         & 0.521  & 0.544  & 0.503  & 0.343  & 0.780   & 0.814  & 0.013  & 0.037   & 0.513   & 0.538  & 0.005  & 0.036   \\ \cline{3-15} 
                       &                                                                                            & Random Patch (Top 3) & -      & -      & -      & -      & 0.780   & 0.821  & 0.044  & 0.017   & 0.521   & 0.545  & 0.013  & 0.021   \\ \hline
\multirow{4}{*}{Fruit} & \multirow{4}{*}{\begin{tabular}[c]{@{}c@{}}Faster R-CNN \\ $\mapsto$ Yolo v5\end{tabular}} & Random Noise         & 0.603  & 0.636  & 0.394  & 0.416  & 0.844   & 0.867  & 0.006  & 0.022   & 0.603   & 0.636  & 0.008  & 0.062   \\ \cline{3-15} 
                       &                                                                                            & Gray Patch           & 0.428  & 0.468  & 0.557  & 0.239  & 0.704   & 0.739  & 0.007  & 0.028   & 0.345   & 0.390  & 0.012  & 0.070   \\ \cline{3-15} 
                       &                                                                                            & Random Patch         & 0.416  & 0.459  & 0.549  & 0.404  & 0.726   & 0.759  & 0.008  & 0.031   & 0.329   & 0.373  & 0.013  & 0.068   \\ \cline{3-15} 
                       &                                                                                            & Random Patch (Top 3) & -      & -      & -      & -      & 0.677   & 0.728  & 0.017  & 0.050   & 0.305   & 0.347  & 0.030  & 0.086   \\ \hline
\end{tabular}%
}
\end{table*}

Overall, adversarial patch attacks in the physical environment achieved average success rates of 69.1, 24.0, and 10.1\% for the Hiding, Creating, and Altering attacks on the snack object detection model. In the fruit object detection model, the rates were 77.6, 44.6, and 10.2\%, respectively. These averages do not always signify meaningful attacks, as they include lower values resulting from robust classes. However, note that certain classes consistently remain vulnerable to adversarial patch attacks. By concretely demonstrating these metrics, we propose that future implementations of automated checkout systems using object detection technology should identify and address these vulnerabilities through simulated attacks, and develop corresponding countermeasures.

Fig. \ref{Fig11} illustrates examples of the evaluation of adversarial patch attacks in our unmanned store testbed. The green borders in each image indicate successful cases, whereas the red borders represent failed cases. We can observe that the patches remain robust even when the size ratio of the patch to the original object, angle at which the patch is viewed, and brightness vary. Particularly, the images of the Hiding attack on the orange object show that the adversarial patch can effectively hide the orange object even when exposed to light or when it is marginally tilted. Moreover, as observed in certain cases of the Altering attack, class manipulation can be achieved while maintaining the size of the bounding box, even if the patch is smaller than the original object. 

However, the failure cases in the first columns of the Creating and Altering attacks demonstrate instances in which the object was created or transformed into an unintended class. These observations suggest that, even if the intended target class attack of an attacker fails, adversarial patches can still disrupt the inference process of the object detection model.

\begin{table*}[t!]
\centering
\caption{Shadow Attacks of Adversarial Patches in a Black-box Scenario}
\label{Table06}
\renewcommand{\arraystretch}{1.5}
\resizebox{\textwidth}{!}{%
\begin{tabular}{ccccccccccccccc}
\hline
                       &                                                                                               &                      & \multicolumn{4}{c}{Hiding Attack} & \multicolumn{4}{c}{Creating Attack} & \multicolumn{4}{c}{Altering Attack} \\ \hline
Dataset                & Model                                                                                         & Attack               & $mAP$  & $mAR$  & $CM$   & $CIoU$ & $mAP$   & $mAR$  & $CM$   & $CIoU$  & $mAP$   & $mAR$  & $CM$   & $CIoU$  \\ \hline
\multirow{4}{*}{Snack} & \multirow{4}{*}{\begin{tabular}[c]{@{}c@{}}Yolo v5 (X) \\ $\mapsto$ YOLO v5 (L)\end{tabular}} & Random Noise         & 0.659  & 0.682  & 0.336  & 0.272  & 0.856   & 0.873  & 0.003  & 0.020   & 0.659   & 0.682  & 0.001  & 0.016   \\ \cline{3-15} 
                       &                                                                                               & Gray Patch           & 0.317  & 0.341  & 0.651  & 0.564  & 0.631   & 0.661  & 0.116  & 0.062   & 0.443   & 0.478  & 0.058  & 0.105   \\ \cline{3-15} 
                       &                                                                                               & Random Patch         & 0.289  & 0.313  & 0.717  & 0.576  & 0.629   & 0.657  & 0.112  & 0.068   & 0.456   & 0.491  & 0.046  & 0.095   \\ \cline{3-15} 
                       &                                                                                               & Random Patch (Top 3) & -      & -      & -      & -      & 0.552   & 0.585  & 0.329  & 0.186   & 0.366   & 0.410  & 0.136  & 0.211   \\ \hline
\multirow{4}{*}{Fruit} & \multirow{4}{*}{\begin{tabular}[c]{@{}c@{}}Yolo v5 (X) \\ $\mapsto$ Yolo v5 (L)\end{tabular}} & Random Noise         & 0.603  & 0.636  & 0.394  & 0.416  & 0.844   & 0.867  & 0.006  & 0.022   & 0.603   & 0.636  & 0.008  & 0.062   \\ \cline{3-15} 
                       &                                                                                               & Gray Patch           & 0.308  & 0.342       & 0.650  & 0.543  & 0.644  & 0.700  & 0.206  & 0.075  & 0.309  & 0.359  & 0.092  & 0.135       \\ \cline{3-15} 
                       &                                                                                               & Random Patch         & 0.337  & 0.372   & 0.618  & 0.501  & 0.642  & 0.704  & 0.210  & 0.075  & 0.320  & 0.367  & 0.104  & 0.143   \\ \cline{3-15} 
                       &                                                                                               & Random Patch (Top 3) & -      & -      & -      & -      & 0.611  & 0.691  & 0.452  & 0.155  & 0.293  & 0.343  & 0.256  & 0.205   \\ \hline
\end{tabular}%
}
\end{table*}

\subsection{Adversarial Patch Attacks in Black-Box Scenario}\label{subsec5_3}
\subsubsection{Model Transfer Attack}\label{subsec5_3_1}
We analyzed the effectiveness of adversarial patch attacks in a black-box setting. In these scenarios, attackers typically attempt a model transfer attack\cite{ma2023transferable}. This approach involves an attacker building an arbitrary surrogate model, training adversarial patches on it, and then querying the victim model. We assumed YOLO v5 was the victim model for detecting snacks and fruits in a digital environment and used Faster R-CNN as the surrogate model for training adversarial patches.

Table \ref{Table05} lists the experimental results. All three types of attacks achieve improved $CM$ and $CIoU$ values over random noise; however, the $CM$ values indicate that the Creating and Altering attacks achieve only approximately 1\% misdetection performance in $CM$. Furthermore, the attack performances of the top-three target classes with the highest attack success rates remain at a maximum of 4.4\%. In conjunction with Table \ref{Table03}, we observed that the adversarial patch attacks were more challenging on Faster R-CNN than on YOLO v5. We hypothesize that this is because of the relatively greater difficulty experienced by attackers when observing the training data distribution in a Faster R-CNN to identify meaningful adversarial perturbations, making it challenging to create effective adversarial patches for transfer attacks. Overall, while transfer attacks for adversarial patches in Creating and Altering attacks are realistically very difficult tasks, a Hiding attack can still pose a threat to fully unmanned stores based on a $CM$ of up to 55.7\%.

\subsubsection{Shadow Attack}\label{subsec5_3_2}
We also evaluated a scenario in which an attacker, assuming that they can repeatedly query the victim object detection model to obtain inference results, constructs a shadow model to generate adversarial patches without having detailed knowledge of the model. The attacker treats the bounding box predictions from the victim model on the input images as the ground truth to form an arbitrary dataset. Subsequently, following a model extraction process, the attacker trains their model, which has a different architecture from that of the victim model, on this dataset. This approach is a representative method for approximating the decision boundaries of a victim model in a black-box environment, thereby alleviating the difficulty of an attack \cite{zhang2022generating}. The attacker then uses the trained shadow model to generate adversarial patches.

We set YOLO v5 (X) as the shadow model and used YOLO v5 (L) as the victim model. In Table \ref{Table06}, we can observe a clear improvement in the attack success rates for all attacks as compared to model transfer attack. The $CM$ values of the Hiding attack are as high as 0.717 and 0.650, and the $CIoU$ values are 0.576 and 0.543 for the snack and fruit datasets, respectively. Moreover, the Creating attack improved the $CM$ to 0.452 for the fruit object detection model in the top-three target classes, and the Altering attack recorded a $CM$ of 0.256. They suggest that repeated queries and exposure to inference result in an automated checkout system that could potentially evolve into a substantial digital security threat.

\begin{figure}[t!]
\centering
\includegraphics[width=0.48\textwidth]{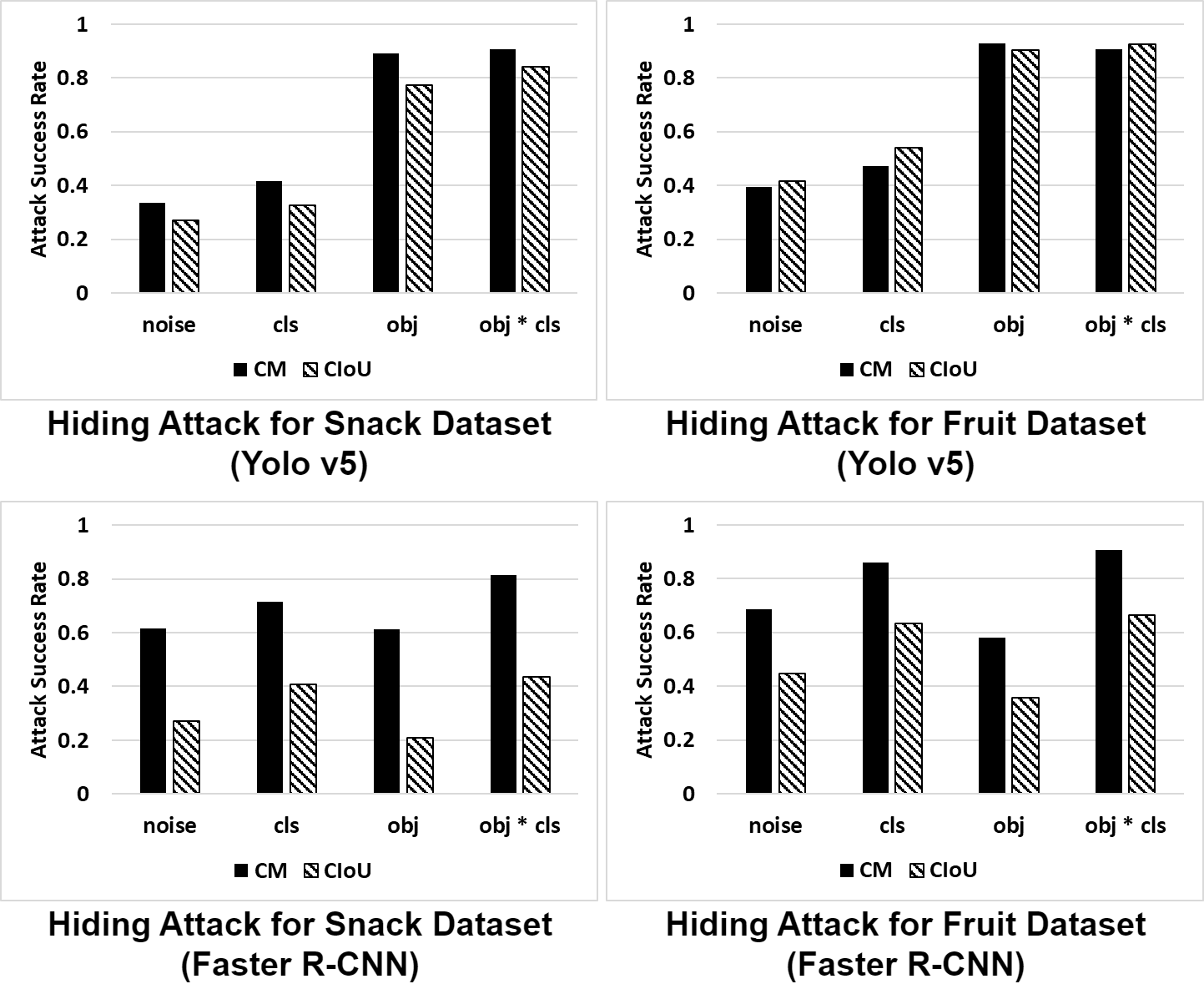}
\caption{Performance differences in Hiding attack based on the settings of the adversarial detection loss function in each experimental case.}
\label{Fig19}
\end{figure}

\subsection{Performance Changes According to Loss Function Settings}\label{subsec5_4}

\subsubsection{Settings of Adversarial Detection Loss Function in Hiding Attack}\label{subsec5_4_1}
In Section \ref{subsubsec3_4_1}, we define (\ref{formula12}), which outlines three cases for setting the adversarial detection loss function $L_{adv}$ for a Hiding attack. We observed the performance differences across the three cases. Fig. \ref{Fig19} shows the $CM$ and $CIoU$ in each case. On the x-axis of each graph, ``noise'' represents the random noise-based occlusion patch, which generally shows lower attack performance as compared to those of the intentionally trained adversarial patches. We observed that simultaneously minimizing both $y_{obj}$ and $y_{cls}$ yielded the highest attack effectiveness in all cases.

\begin{figure}[t!]
\centering
\includegraphics[width=0.48\textwidth]{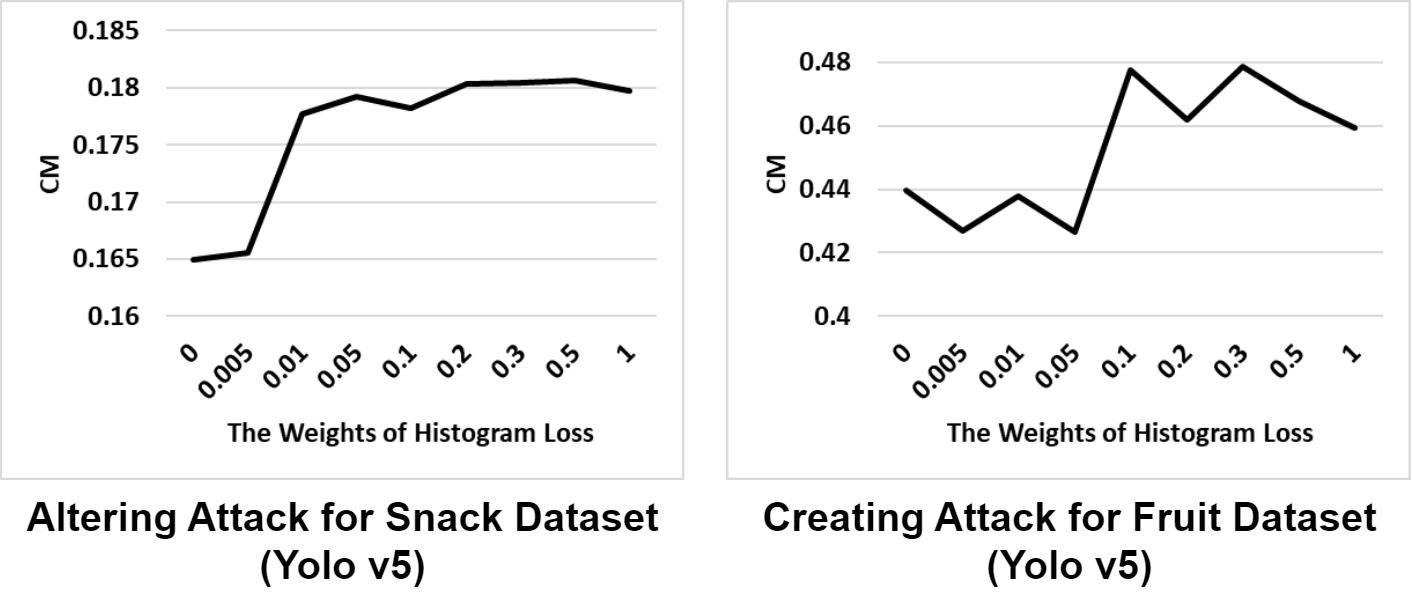}
\caption{Changes in attack success rates in the Creating and Altering attacks with different weights of color histogram similarity loss.}
\label{Fig20}
\end{figure}

An interesting observation is that the YOLO v5 model effectively learns adversarial patches when the objective function targets $y_{obj}$, whereas the Faster R-CNN model is effective when targeting $y_{cls}$. This difference can be inferred from the detection mechanisms of the two models. YOLO v5 predicted multiple bounding boxes for each grid cell along with the corresponding $y_{obj}$ and $y_{cls}$ values. In the early stages of the YOLO algorithm, $y_{obj}$ was used to determine the presence of an object. Thus, if the model strongly recognizes the existence of an object, manipulating $y_{cls}$ alone may not be sufficient for hiding the object. Conversely, Faster R-CNN uses an RPN to pre-identify the likelihood of the presence of an object and then classifies the object within each region. In this two-step process, the RPN, which is highly optimized, plays a crucial role in determining the likelihood of an object within a specific region. Therefore, simply minimizing $y_{obj}$ may not be effective in hiding an object because of the additional analysis stage that determines the class of the object. In conclusion, the vulnerabilities of the object detection models identified through our extensive experiments can be attributed to these structural factors. The insights we uncovered is valuable for building more robust models in the future.

\subsubsection{Weights of Histogram Loss}\label{subsec5_4_2}
In this section, we describe how the attack success rate $CM$ of adversarial patches changes with different weights $\lambda_{His}$ for our proposed loss function $L_{His}$. Fig. \ref{Fig20} shows $CM$ as $\lambda_{His}$ increases from 0 to 1, indicating that $L_{His}$ enhances the effectiveness of both the Creating and Altering attacks. For the Altering attack on the snack object detection model (left side of the figure), $CM$ increases from 0.165 to a maximum of 0.180. For the Creating attack on the fruit object detection model (right side of the figure), $CM$ increases from 0.440 to a maximum of 0.480. In our experimental setup, $\lambda_{His}=0.3$ was the most effective; thus, this value was selected for our main experiments. In other environments, attackers may need to empirically determine the optimal value.

\section{Discussion}\label{sec6}
\subsection{Challenges Faced by the Attacker}\label{subsec6_1}
\subsubsection{Physical Adversarial Patch}
Several previous studies \cite{hu2021naturalistic,doan2022tnt,zhu2023tpatch} have explored the creation of natural adversarial patches to evade human detection by mimicking natural objects. Unlike these studies, our adversarial patches did not prioritize image naturalness, potentially limiting their effectiveness in physical settings. However, integrating a semantic objective function to ensure naturalness reduces the noise subspace size, making the attack more challenging \cite{godfrey2023many}. In our experiments, Creating and Altering attacks showed unsatisfactory success rates across target classes, highlighting the difficulty of generating effective natural adversarial patches.

Meanwhile, adversarial patch attacks can still pose significant threats despite their low overall success rates. Even if not consistently activated in all frames, malfunctions in specific frames can disrupt store operations, causing errors in promotions, product identification, and placement. This leads to inefficiencies as additional employee time and resources are required to fix these issues. Persistent detection errors can also harm customer satisfaction, trust in automated systems, and the store's reputation, while exposing security vulnerabilities. Both attackers and security managers must recognize that such attacks, despite their low success rates, can have cumulative impacts that significantly affect the viability and profitability of unmanned stores.

\subsubsection{Class-agnostic Attack}\label{subsubsec6_1_2}
We observed that the performances of the Creating and Altering attacks varied significantly depending on the target class. Moreover, we observed that the same adversarial patch could produce different attack effects depending on the original class, as shown in Figs. \ref{Fig10} and \ref{Fig09}. We inferred that this phenomenon is due to certain non-robust features\cite{kim2021distilling} in the dataset that was used to train the object detection model. These features are highly sensitive to small transformations, such as rotation, translation, and noise, and can be considered as artifacts of the training data and model architecture that do not align with human perception. In environments such as unmanned stores, where the color and shape of data within each class are consistent, a causal graph must be constructed for external factors\cite{yuanxin2023fca} such as object and image features, to explore the key factors affecting non-robust features, which will be considered in our future work. Furthermore, in future research, we aim to explore scenarios in which attackers can maximize the feasibility of attacks by pre-observing the robustness of the desired target object through these factors.

\subsection{Challenges Faced by the Defender}\label{subsec6_2}
Many researchers have proposed countermeasures against adversarial patch attacks on object detection models, categorized into three types. Among these, \textbf{detection and removal} methods \cite{bunzel2023adversarial, rossolini2023defending, tarchoun2023jedi} and \textbf{certifiably robust defense} frameworks \cite{xiang2023objectseeker, xiang2022patchcleanser, xiang2021patchguard++} each involve high computational and time costs, which can limit the number of frames processed. Additionally, these approaches are often primarily focused on Hiding attacks and face challenges in dealing with multiple patches simultaneously, thus being restrictive for real-time applications.

Conversely, robust training \cite{rao2020adversarial, gittings2020vax} aims to enhance model robustness during the training phase, potentially valuable for fully unmanned stores without real-time inference delays. However, this approach is under-researched and struggles to provide robustness against unseen adversarial patches, often focusing on specific patterns of adversarial patches. In addition, the tradeoff between clean performance and adversarial robustness must be explored.

Improving the robustness of real-time object detection remains a significant challenge. Strengthening the boundaries of $y_{cls}$ and $y_{obj}$ during adversarial patch training, as suggested in Section \ref{subsec5_4_1}, is a promising approach. Analyzing and combining the output processes of YOLO series and Faster R-CNN architectures could lead to new solutions, which we leave for future research.

\section{Conclusion}\label{sec7}
This study addresses the security vulnerabilities posed by adversarial patch attacks in fully unmanned stores using AI-based object detection technology. We investigated three primary types of attacks: Hiding, Creating, and Altering attacks, and analyzed their effects through extensive experiments. Our results demonstrated that adversarial patches can severely disrupt the operation of unmanned stores. Notably, we proposed a novel objective function, the color histogram similarity loss, which leverages the color information of the target class object to enhance attack success rates. Starting with attacks in a digital environment, we evaluated the effectiveness of physical attacks in a testbed that simulated an actual unmanned store. Furthermore, we demonstrated that even in a black-box scenario, where direct access to model parameters is unavailable, shadow attacks can significantly improve attack success rates.

To the best of our knowledge, studies on AI security vulnerabilities in fully unmanned stores are scarce. Our study provides essential information for the safe deployment of AI in real-world retail environments. Moreover, our discussion offers insights into the generation of more threatening adversarial patches and the exploration of effective countermeasures. Finally, we aim to design optimal object detection models that are robust against adversarial patch attacks while maintaining the efficiency of fully unmanned stores.

\ifCLASSOPTIONcaptionsoff
  \newpage
\fi

\bibliographystyle{IEEEtran}
\bibliography{Reference}

\vfill
\end{document}